\documentstyle[preprint,aps]{revtex}

\begin{document}

\draft


\tighten

\title{Two-dimensional Chiral Anomaly in Differential Regularization}

\author{W.F. Chen\renewcommand{\thefootnote}{\dagger}\footnote{E-mail:
wchen@theory.uwinnipeg.ca}}
 
\address{Department of Physics, 
 University of Winnipeg, Winnipeg, Manitoba, R3B 2E9 Canada \\
 and \\
Winnipeg Institute for Theoretical Physics, Winnipeg, Manitoba}

\maketitle

\begin{abstract}
 The two-dimensional chiral anomaly is calculated using
differential regularization. It is shown that the anomaly emerges
naturally in the vector and axial Ward identities on the same footing
as the four-dimensional case. The vector gauge symmetry can be achieved 
by an appropriate choice of the  mass scales without introducing
the seagull term. We have analyzed the reason why such a universal 
result can be obtained in differential regularization. 

\vspace{3ex}
 
\end{abstract}

\vspace{3ex}

One of the most remarkable dynamical phenomena in two-dimensional 
massless Quantum Electrodynamics ($QED_2$) is that the photon field 
becomes massive through a dynamical Higgs mechanism induced by the
fermion$\cite{ref1}$. The essence for its occurrence lies in the spontaneous
breaking of the global chiral symmetry $U(1){\times}\widetilde{U}(1)$,
$U(1)$ and $\widetilde{U}(1)$ denoting the Abelian group associated with
the charge and chirality of the fermion, respectively. 
This fact makes $QED_2$ served as a laboratory for understanding the 
vacuum structure of $QCD_4$$\cite{ref3}$. 

The two-dimensional chiral anomaly plays a crucial role in revealing above
dynamical phenomena$\cite{ref2,ref4}$. In fact, the two-dimensional 
chiral anomaly was first found by Johnson$\cite{ref5}$ prior to the discovery
of the famous ABJ chiral anomaly in four dimensions$\cite{ref5a1}$.
Later an explicit perturbative calculation
was also carried out in Ref.$\cite{ref5a2}$. 
In the last decade this anomaly has been re-investigated by various 
perturbative and non-perturbative calculation techniques such as 
dimensional, Pauli-Villars and zeta function regularization schemes
as well as a dispersion relation approach$\cite{ref6}$ and 
topological analysis$\cite{ref7}$. Recently, 
it is calculated again by a quite
old method called analytic regularization$\cite{ref7a1}$. 
However, we think that it is still worthwhile to use  
a newly developed quantum field theory methods to gain insight. This may not 
only reveal some of the new features of the two-dimensional anomaly, 
but also test the feasibility of the modern computation techniques.

In view of the above considerations, 
in this letter we shall use a relatively new 
regularization scheme called differential regularization$\cite{ref8}$
 to calculate the two-dimensional chiral anomaly. The basic idea of this
regularization is quite simple. It works in coordinate space for  
Euclidean field theory and is based on the observation that the higher
order amplitude cannot have a Fourier transform into momentum space
due to the short-distance singularity. Thus one can regulate such an 
amplitude by first writing its singular parts as the derivatives of
the less singular functions, which have a well defined Fourier
transform, and then performing Fourier transformation in partial 
integration and discarding the surface term. In this way one can directly
get a renormalized result.  Up to now this method and its modified
version have been applied successfully to almost every field theories, 
including the supersymmetric ones $\cite{ref8,ref9,ref11,ref12,ref14,ref14a0}$. 
 
The main motivation for us to choose this regularization
scheme to study the two-dimensional chiral anomaly is its great 
advantage over other regularization schemes in preserving
gauge symmetry. The maintenance of the gauge symmetry in differential
regularization is  achieved by appropriately choosing the 
indefinite renormalization scales at the final stage of the calculation.
Therefore, this regularization method can lead to a most democratic
expression for the chiral anomaly. On the other hand, in Pauli-Villars,
dimensional and Zeta function regularization,  
 a renormalization condition is automatically  imposed on the
vector and axial Ward identities when calculating
the chiral anomaly. That is, these regularization schemes themselves
are compatible with the conservation of vector gauge symmetry
and violate the axial vector symmetry, thus one can only
observe an anomalous axial vector Ward identity. Differential 
regularization is different. The study of four-dimensional chiral anomaly 
in terms of differential regularization was initially carried out in 
Ref.$\cite{ref8}$ and $\cite{ref11}$, and it was shown that the anomalous
term can naturally arise in the VWI and AWI on the same footing. Furthermore,
combined with the conformal symmetry of the correlate function,  as initially
proposed by Baker and Johnson$\cite{ref14a1}$, differential regularization
was also applied to investigate the possible two-loop radiative correction to
the chiral anomaly and the axial gravitational anomaly$\cite{ref14a2}$. Therefore, the investigation 
of the chiral anomaly in terms of differential regularization will definitely  
be more powerful in revealing the quantum structure
of the theory than other regularization schemes. In addition, 
differential regularization does not
modify the original Lagrangian at all, 
hence it never shifts the value of a primitively divergent Feynman
diagram away from its singularities, and neither does it extend
the dimensionality of the space-time or introduce a regulator$\cite{ref11}$.
Thus this regularization method is free from the ambiguity in defining
the dimensional continuation of $\gamma_5$, which can occur in dimensional 
regularization. The details of the calculation in the differential 
regularization are also simpler than in Pauli-Villars and Zeta function regularization.

We now set up the framework for calculating this anomaly. The Lagrangian 
of $QED_2$ with one flavour fermion in Euclidean space is read 
as follows$\cite{ref4}$: 
\begin{eqnarray}
{\cal L}=\bar{\psi} (i/\hspace{-2mm}{\partial}-e/\hspace{-2.5mm}A-m)\psi
-\frac{1}{4}F_{\mu\nu}F_{\mu\nu}, ~~~~~~\mu, \nu=1,2,
\label{eq1}
\end{eqnarray} 
where the $\gamma$-matrices are chosen as the two-component form
\begin{eqnarray}
\gamma_1=\sigma_2, ~~\gamma_2=-\sigma_1, ~~\gamma_5=-i\gamma_1\gamma_2=
\sigma_3.
\end{eqnarray}
Classically, the following vector and axial currents 
\begin{eqnarray}
j_{\mu}=\bar{\psi}\gamma_\mu\psi, ~~j_{\mu}^5=i\bar{\psi}\gamma_\mu\gamma_5\psi
\end{eqnarray}
satisfy the relations
\begin{eqnarray}
\partial_\mu j_\mu=0, ~~\partial_\mu j_\mu^5=2im j_5
\end{eqnarray}
with $j_5$ being the pseudo-scalar current, $j_5{\equiv}\bar{\psi}\gamma_5\psi$.

 In $QED_2$ the chiral anomaly  comes from the
two-point function$\cite{ref7}$
\begin{eqnarray}
\Pi_{\mu\nu}^5(x-y)=\langle T[j_{\mu}(x)j^5_{\nu}(y)]\rangle.
\label{eq12}
\end{eqnarray}
This is contrary to the four-dimensional case, where the chiral
anomaly comes from the triangle composed of the axial and vector 
currents. Due to the explicit relation     
$i\gamma_\mu\gamma_5=\epsilon_{\mu\nu}\gamma_\nu$, $T_{\mu\nu}^5$ is 
relevant to the vacuum polarization tensor
\begin{eqnarray}
\Pi_{\mu\nu}^5(x-y)=\epsilon_{\nu\rho}\Pi_{\rho\mu}(x-y).
\label{eq9}
\end{eqnarray}
Thus $\Pi_{\mu\nu}^5$ can be calculated via the vacuum polarization tensor.

If the classical symmetries of the theory were preserved at the quantum level
there would exist the following vector Ward identity (VWI),
\begin{eqnarray}
\partial_\mu^x\Pi_{\mu\nu}^5(x-y)=\partial_{\mu}^x\Pi_{\mu\nu}(x-y)
=\partial_{\nu}^y\Pi_{\mu\nu}(x-y)=0
\end{eqnarray}
and the corresponding axial vector one (AWI),
\begin{eqnarray}
\partial_\nu^y\Pi_{\mu\nu}^5(x-y)=2im \Pi_{\mu}^5 (x-y)
\end{eqnarray}
with that
$\Pi_{\mu}^5 (x-y){\equiv}\langle T[j_{\mu}(x)j_5(y)]\rangle$.
Later it can be seen that the above Ward identities will
be violated due to the chiral anomaly.

 For massless $QED_2$, the propagator of the fermion in Euclidean 
space is
\begin{eqnarray}
S(x)=-\frac{1}{2\pi}/\hspace{-2.3mm}\partial\ln\frac{1}{x}, 
\label{eq15}
\end{eqnarray} 
where and later we denote $x{\equiv}|x|$; While for the massive case, 
we have
\begin{eqnarray}
S(x)=\frac{1}{2\pi}(/\hspace{-2.3mm}\partial-m) K_0(mx),
\end{eqnarray}
where $K$ is the  modified Bessel function of the second kind. 

Let us first consider the massless case. 
With the propagator (\ref{eq15}) we write down the vacuum polarization 
tensor
\begin{eqnarray}
\Pi_{\mu\nu}(x)&=&-e^2\mbox{Tr}\left[\gamma_\mu S(x)\gamma_\nu S(-x)\right]
=\frac{e^2}{4\pi^2}\mbox{Tr}(\gamma_\mu\gamma_\alpha\gamma_\nu\gamma_\beta)
\frac{x_\alpha x_\beta}{x^4}\nonumber\\
&=&\frac{e^2}{2\pi^2}\left(
\frac{2x_\mu x_\nu}{x^4}-\frac{\delta_{\mu\nu}}{x^2}\right),
\label{eq18}
\end{eqnarray}
where we have used the relation $\partial_\mu f(x)={x_\mu}/{x}[d/dxf(x)]$
and the two-dimensional $\gamma$-matrix trace formula $\mbox{Tr}(\gamma_\mu\gamma_\alpha\gamma_\nu\gamma_\beta)
=2\left(\delta_{\mu\alpha}\delta_{\nu\beta}
-\delta_{\mu\nu}\delta_{\alpha\beta}
+\delta_{\mu\beta}\delta_{\nu\alpha}\right)$.

The term with the tensor structure $x_\mu x_\nu$ can be rewritten as
\begin{eqnarray}
\frac{x_\mu x_\nu}{x^4}=\frac{1}{2}\left(\partial_\mu \partial_\nu
\ln\frac{1}{x}+\frac{\delta_{\mu\nu}}{x^2}\right).
\label{eq21}
\end{eqnarray}
Upon substitution of (\ref{eq21}) into the vacuum polarization tensor
(\ref{eq18}), it seems that the term with the tensor structure 
$\delta_{\mu\nu}$ could cancel. In fact, this is not allowed since
in two dimension the term $\sim 1/x^2$  is obviously singular. It is 
analogous to the fact that two divergent terms with the same form 
but opposite sign in momentum
space cannot be canceled, only after a regularization procedure is
implemented so that they become well defined, the substraction 
operation can work safely. Otherwise, a finite term will probably 
lost since the difference of two infinite quantities is generally 
not zero. We therefore use differential regularization schemes 
to make $1/x^2$ well defined. Writing
\begin{eqnarray}
\frac{1}{x^2}=\partial^2 f(x)=\frac{1}{x}\frac{d}{dx}
\left[x \frac{d}{dx}f (x)\right],
\end{eqnarray} 
we get 
\begin{eqnarray} 
f(x)=-\frac{1}{2}\ln (M x)\ln\frac{1}{x},
\end{eqnarray}
and hence the regulated version of $1/x^2$ is 
\begin{eqnarray}
\left(\frac{1}{x^2}\right)_R=-\frac{1}{2}\partial^2\left[\ln (Mx)\ln \frac{1}{x}\right], 
\label{eq24}
\end{eqnarray}
where $M$ can be explained as 
the renormalization scale, which is necessary for the
differential regularization. (\ref{eq24}) are exactly the analogue 
of four-dimensional case$\cite{ref8}$, 
$1/x^4=-1/4\partial^2 [\ln (M^2x^2)/x^2]$. 

With (\ref{eq21}) and (\ref{eq24}) the differential regulated 
 vacuum polarization tensor is obtained
\begin{eqnarray}
\Pi_{\mu\nu}(x)&=&\frac{e^2}{2\pi^2}\left\{\partial_\mu\partial_\nu
\ln\frac{1}{x}-\delta_{\mu\nu}\frac{1}{2}\partial^2\left[\ln (M_1x)
\ln \frac{1}{x}\right]+\delta_{\mu\nu}\frac{1}{2}\partial^2\left[\ln (M_2x)
\ln \frac{1}{x}\right]\right\}\nonumber\\
&=&\frac{e^2}{2\pi^2}\left(\partial_\mu\partial_\nu
-\delta_{\mu\nu}\frac{1}{2}\ln\frac{M_1}{M_2}\partial^2\right)
\ln\frac{1}{x}.
\label{eq25}
\end{eqnarray}
Eq.(\ref{eq9}) gives the two-point function between the vector and 
axial vector currents
\begin{eqnarray}
\Pi^5_{\mu\nu}(x)=\epsilon_{\nu\rho}\Pi_{\rho\mu}(x)
=\frac{e^2}{2\pi^2}\epsilon_{\nu\rho}\left(\partial_\rho\partial_\mu
-\delta_{\rho\mu}\frac{1}{2}\partial^2\ln\frac{M_1}{M_2}\right)
\ln \frac{1}{x}.
\label{eq26}
\end{eqnarray}

Just like the four-dimensional case, the results 
(\ref{eq25}) and (\ref{eq26}) have displayed 
a most democratic expression for the two-dimensional chiral anomaly. 
With a general choice  
\begin{eqnarray}
M_1=e^n M_2,
\label{eq34}
\end{eqnarray}
we obtain the two-point functions 
\begin{eqnarray}
\Pi_{\mu\nu}(x)&=&\frac{e^2}{2\pi^2}\left(\partial_\mu\partial_\nu
-\frac{n}{2}\delta_{\mu\nu}\partial^2\right)\ln\frac{1}{x},\nonumber\\
\Pi_{\mu\nu}^5(x)&=&\frac{e^2}{2\pi^2}\epsilon_{\nu\rho}
\left(\partial_\rho\partial_\mu-\frac{n}{2}\delta_{\rho\mu}\partial^2\right)
\ln\frac{1}{x}.
\label{eq}
\end{eqnarray} 
and the corresponding vector and axial vector Ward identities,
\begin{eqnarray}
\partial_\mu\Pi^5_{\mu\nu}(x)&=&\left(\frac{n}{2}-1\right)
\frac{e^2}{\pi}\epsilon_{\nu\mu}\partial_\mu\delta^{(2)}(x), \nonumber\\
\partial_\nu\Pi^5_{\mu\nu}(x)&=&\frac{n}{2}\frac{e^2}{\pi}
\epsilon_{\nu\mu}\partial_\nu\delta^{(2)}(x).
\end{eqnarray}

The choice  $n=2$ is the case we are familiar with, where the
VWI is satisfied, while AWI becomes anomalous,
\begin{eqnarray}
\partial_\mu\Pi_{\mu\nu}(x)&=&\partial_\nu\Pi_{\mu\nu}(x)
=\partial_\mu\Pi^5_{\mu\nu}(x)=0,\nonumber\\
\partial_\nu\Pi^5_{\mu\nu}(x)&=&-\frac{e^2}{2\pi}
\epsilon_{\nu\mu}\partial_\nu\partial^2\ln\frac{1}{x}
=\frac{e^2}{\pi}\epsilon_{\nu\mu}\partial_{\nu}\delta^{(2)}(x).
\label{eq31}
\end{eqnarray}
The $n=0$ choice just corresponds to the conservation of axial vector 
current but an anomalous vector current. In any choices, both of the vector 
and axial Ward identities cannot be fulfilled simultaneously.

 The massive case can be  calculated in a similar way. 
The vacuum polarization tensor is 
\begin{eqnarray}
\Pi_{\mu\nu}(x)&=&-\frac{e^2}{4\pi^2}\mbox{Tr}\left[\gamma_\mu
\left(/\hspace{-2mm}\partial-m\right)K_0(mx)
\gamma_\nu\left(-/\hspace{-2mm}\partial-m\right)K_0(m x)\right]\nonumber\\
&=&\frac{e^2}{4\pi^2}\left\{\mbox{Tr}\left(\gamma_\mu\gamma_\alpha
\gamma_\nu\gamma_\beta\right)x_\alpha
x_\beta\left[\frac{1}{x}\frac{d}{dx}K_0(mx)\right]^2
-m^2\mbox{Tr}(\gamma_\mu\gamma_\nu)\left[K_0(mx)\right]^2\right\}\nonumber\\
&=&\frac{e^2}{2\pi^2}\left[2x_\mu x_\nu \left(\frac{m K_1(mx)}{x}\right)^2
- m^2\delta_{\mu\nu}\left(\left[K_0(mx)\right]^2+\left[K_1(mx)\right]^2
\right)\right].
\label{eq37}
\end{eqnarray} 
Making use of the following two-dimensional differential operations,
\begin{eqnarray}
2x_\mu x_\nu \frac{m^2\left[K_1(mx)\right]^2}{x^2}
&=&\partial_\mu\partial_\nu\left[m^2 x^2\left([K_0(mx)]^2-[K_1(mx)]^2\right)
+mxK_0(mx)K_1(m x)\right]\nonumber\\
&&-\delta_{\mu\nu}m^2\left([K_0(mx)]^2-[K_1(mx)]^2\right); \nonumber\\
2m^2[K_0(mx)]^2&=&\partial^2\left[m^2 x^2\left([K_0(mx)]^2-[K_1(mx)]^2\right)
+mxK_0(mx)K_1(m x)\right],
\end{eqnarray}
we can write the vacuum polarization tensor of the massive $QED_2$ 
as the following form,
\begin{eqnarray}
\Pi_{\mu\nu}(x)&=&\frac{e^2}{2\pi^2}\left\{\partial_\mu\partial_\nu
\left[m^2 x^2 [K_0(mx)]^2-m^2 x^2 [K_1(mx)]^2+mx K_0(mx)K_1(mx)\right]\right.
\nonumber\\
&&\left.-\delta_{\mu\nu} m^2\left([K_0(mx)]^2-[K_1(mx)]^2\right)
-\delta_{\mu\nu} m^2\left([K_0(mx)]^2+[K_1(mx)]^2\right)\right\}
\nonumber\\
&=&\frac{e^2}{2\pi^2}\left\{\left(\partial_\mu\partial_\nu
-\delta_{\mu\nu}\partial^2\right)
\left[m^2 x^2 \left([K_0(mx)]^2-[K_1(mx)]^2\right)\right.\right.\nonumber\\
&& \left.\left.- mx K_0(mx)K_1(mx)\right]+\delta_{\mu\nu}\left(m^2[K_1(mx)]^2
-m^2[K_1(mx)]^2\right)\right\}.
\label{eq24n}
\end{eqnarray}
The last two $m^2[K_1(mx)]^2$ terms of Eq.(\ref{eq24n}) 
cannot be canceled naively since they are singular as $x{\sim}0$. 
Considering the short-distance expansion of $K_1(mx)$,
\begin{eqnarray}
K_1(mx)\stackrel{x{\sim}0}{\longrightarrow}
\frac{1}{mx}+mx\left[\frac{1}{2}\ln\frac{mx}{2}
+\frac{1}{2}\gamma-\frac{1}{4}\right]+{\cal O}(x^2)
\end{eqnarray}
with $\gamma$ being the Euler constant, we can see that 
in Eq.(\ref{eq24n}) the short-distance singularity is only carried by
the leading term $1/x^2$, the other terms are finite and they are exactly
canceled. Therefore, employing Eq.(\ref{eq24}) again, we finally get
the differential regulated form of the vacuum polarization tensor for the
massive case,
\begin{eqnarray}
\Pi_{\mu\nu}(x)&=&\frac{e^2}{2\pi^2}\left\{\left(
\partial_\mu\partial_\nu-\delta_{\mu\nu}\partial^2\right)
\left[m^2 x^2 \left([K_0(mx)]^2- [K_1(mx)]^2\right)\right.\right.\nonumber\\
&& \left.\left.- mx K_0(mx)K_1(mx)\right]+\delta_{\mu\nu}\left(
\frac{1}{x^2}-\frac{1}{x^2}\right)\right\}\nonumber\\
&=&\frac{e^2}{2\pi^2}\left\{\left(\partial_\mu\partial_\nu
-\delta_{\mu\nu}\partial^2\right)
\left[m^2 x^2 \left([K_0(mx)]^2-[K_1(mx)]^2\right)\right.\right.\nonumber\\
&& \left.\left.- mx K_0(mx)K_1(mx)\right]+\delta_{\mu\nu}
\frac{1}{2}\ln\frac{M_1}{M_2}\partial^2\ln\frac{1}{x}\right\}.
\label{eq26n}
\end{eqnarray}
To check the Ward identities of the massive case, we 
need the two-point function (\ref{eq12}),
\begin{eqnarray}
\Pi_\mu^5(x)&=&-\frac{e^2}{4\pi^2}\mbox{Tr}\left[\gamma_\mu
\left(/\hspace{-2mm}\partial-m\right)K_0(mx)
\gamma_5\left(-/\hspace{-2mm}\partial-m\right)K_0(mx|)\right]\nonumber\\
&=&\frac{ie^2}{\pi^2}m\epsilon_{\mu\nu}K_0(mx)\partial_\nu K_0(mx)
=\frac{ie^2}{2\pi^2}m\epsilon_{\mu\nu}\partial_\nu\left[K_0(mx)\right]^2.
\label{eq46}
\end{eqnarray}
In order to satisfy the VWI, we must impose $M_1=M_2$. Then using
the identity for the two-dimensional
Euclidean scalar propagator, 
\begin{eqnarray}
\partial^2 K_0(mx)=m^2 K_0(mx)-2\pi \delta^{(2)}(x).
\end{eqnarray}
one can check that the VWI is satisfied and the AWI 
becomes anomalous,
\begin{eqnarray}
\partial_{\mu}\Pi_{\mu\nu}(x)=\partial_{\nu}\Pi_{\mu\nu}(x)
=\partial_{\mu}\Pi_{\mu\nu}^5(x)=0; \nonumber\\
\partial_{\nu}\Pi_{\mu\nu}^5=2i  m \Pi_{\mu}^5(x)
+\frac{e^2}{\pi}\epsilon_{\nu\mu}\partial_\nu\delta^{(2)}(x).
\end{eqnarray}
On the other hand, if we choose $M_1=e^2M_2$, the AWI will be satisfied
while the VWI is violated. Note that the mass scale choices for carrying 
the Ward identities out in the massive case are different from those in the massless case. 

Eqs. (\ref{eq25}), (\ref{eq26}), (\ref{eq26n}) and the calculation 
deriving them show that the origin of the two-dimensional anomaly 
in the differential regularization is the same as in the
four-dimensional case$\cite{ref11}$. To regulate 
two different singular pieces, one has
to introduce two different mass scales. The terms with these two mass scales
actually can combine into a finite amplitude with a continuous 
one-parameter shift degree of freedom given by the quotient of these two scales.
This degree of freedom cannot accommodate  the vector
and axial vector Ward identities simultaneously so that the 
two-dimensional chiral anomaly has to emerge and manifests itself in 
the vector and axial vector Ward identities on the same footing.  

In summary,  we have obtained two-dimensional anomaly neatly
in terms of differential regularization. In comparison with other approaches
such as dimensional and Pauli-Villars regularization, this regularization
scheme has clearly exhibited the nature of the two-dimensional 
chiral anomaly and especially,
the anomaly has manifested itself in the vector and axial Ward
identities impartially. Furthermore, the impartiality is achieved 
automatically in differential regularization and does not require 
an explicit particular choice of the renormalization conditions on the
physical amplitude as other regularization schemes$\cite{ref15}$. 
This has not only demonstrate the applicability of differential 
regularization to two-dimensional gauge theory, but also has revealed
an intrinsic anomaly structure of the theory.

 Another point which should be emphasized is that we have overcome
the difficulty of the gauge non-invariance of the two-point function
of $QED_2$ in coordinate space. It is known that the vacuum polarization
tensor calculated before in coordinate space is not gauge invariant,
a covariant seagull term has to be added to gain the gauge
 symmetry$\cite{ref16}$. Of course, the reason for this is that
the singularity is not properly regulated. Here in the framework of
differential regularization we have achieved the gauge invariance by an 
appropriate choice of the undetermined mass scales.

The above result obtained in differential regularization is not accidental
and the profound reason lies in the nice features presented by 
differential regularization itself$\cite{ref25}$ 
As it is shown above, the basic operation in differential 
regularization is replacing a singular
term  by the derivative of another less singular function. This operation
has provided a possibility to add arbitrary local terms into
the amplitude. Whenever performing such a operation, we are 
introducing into a new arbitrary local term into the quantum effective
action. For the issue
we are considering, the vacuum polarization tensor in $QED_2$, this 
arbitrariness is parameterized by the ratio of the two renormalization
scales, $M_1/M_2$. In fact, this case is exactly the phenomenon 
emphasized by Jackiw$\cite{ref17}$ recently. According to 
renormalization theory, the introduction of an arbitrary local 
term in the amplitude is equivalent to
the addition of a finite counterterm to the Lagrangian. Therefore,
differential regularization can yield the most
general quantum effective action.
In particular, as stated in the introduction,
this local parameterized ambiguity in differential regularization can 
be fixed at the final stage of the calculation by inputing 
some physical requirements. This special feature of differential 
regularization has formed a sharp contrast to other regularization schemes
such as dimensional, Pauli-Villars and cut-off regularization etc. These
regularization methods, together with given renormalization prescription, 
can fix the arbitrary local terms automatically. 
Therefore, differential regularization can give
a more universal result than any other regularization method, since it does not 
impose any preferred choice on the amplitude at the beginning 
of implementing the regularization. This is just the reason why the
chiral anomaly can emerge in both vector and axial vector Ward identities
on the same footing, while in dimensional and Pauli-Villars regularization 
method, the anomaly only reflects in the axial Ward identity since
the vector gauge symmetry has already been fixed in these regularization
schemes.

\acknowledgments

 This work is supported by the Natural Sciences and Engineering Research
Council of Canada. I am greatly indebted  to Prof. G. Kunstatter for his 
important remarks and  improvements on this manuscript. I would like to 
thank Dr. M. Carrington and Prof. R. Kobes for their encouragements and help. 
I am also grateful to Prof. R. Jackiw for his comments and especially
informing us Ref.$\cite{ref17}$. I am especially obliged to 
Dr. M. Perez-Victoria for his comments and enlightening discussions 
on differential regularization. Last but not the least, I would like to thank
 the referee for his pointing out a big mistake in the first submitted version.

\end{document}